\documentclass[12pt]{iopart}

\def\um{\mu\mbox{m}}

\def\psec{\mbox{ps}}

\def\Wcm2{\mbox{W cm}^{-2}}
\def\Wcmum2{\mbox{Wcm}^{-2}\mu\mbox{m}^{2}}
\def\cm3{\mbox{cm}^{-3}}
\def\Vcm{\mbox{V cm}^{-1}}
\def\MG{\mbox{MGauss}}

\usepackage{iopams}  

\usepackage{graphicx}
\usepackage{color}
\usepackage{epsfig}

\begin{document}

\title
{Ion dynamics and coherent structure formation following laser pulse self-channeling}

\author{Andrea Macchi$^{1,2}$, Alessandra Bigongiari$^2$, Francesco Ceccherini$^2$, Fulvio Cornolti$^2$, Tatiana V Liseikina$^2$\thanks{On leave from the Institute for Computational Technologies, SD-RAS, Novosibirsk, Russia}, 
Marco Borghesi$^3$, Satyabrata Kar$^3$, and Lorenzo Romagnani$^3$}
\address{$^1$polyLab, CNR-INFM, Pisa, Italy}\ead{macchi@df.unipi.it}
\address{$^2$Dipartimento di Fisica ``Enrico Fermi'', Universit\`a di Pisa, Largo Bruno Pontecorvo 3, I-56127 Pisa, Italy}
\address{$^3$School of Mathematics and Physics, the Queen's University of Belfast, Belfast BT7 1NN, UK}

\begin{abstract}
The propagation of a superintense laser pulse in an underdense, inhomogeneous 
plasma has been studied numerically by two-dimensional particle-in-cell 
simulations on a time scale extending up to several picoseconds. The effects
of the ion dynamics following the charge-displacement self-channeling of the 
laser pulse have been addressed. Radial ion acceleration leads to 
the ``breaking'' of the plasma channel walls, causing an inversion
of the radial space-charge field and the filamentation of the laser pulse. 
At later times a number of long-lived, quasi-periodic field structures  
are observed and their dynamics is characterized with high resolution. Inside 
the plasma channel, a pattern of electric and magnetic fields resembling both 
soliton- and vortex-like structures is observed.
\end{abstract}

\pacs{52.38.-r 52.38.Hb 52.38.Kd}

\submitto{\PPCF}

\maketitle

\section{Introduction}
The propagation of superintense laser pulses through low-density plasmas 
gives rise to a variety of nonlinear electromagnetic phenomena 
\cite{askaryanPPCF97,bulanovRPP01}. 
As a general issue the response of the plasma
is nonlinear due both to relativistic effects (hence the definition of 
``relativistic optics'' \cite{mourouRMP06}) and to the intense ponderomotive
force (i.e., the radiation pressure) which strongly modifies the local
plasma density. Probably, the example of such dynamics which has been
mostly investigated is the  self-focusing, channeling and filamentation of 
the laser pulse \cite{sunPF87,moriPRL88,borisovPRA92,monotPRL95,borghesiPRL97,krushelnickPRL97,sarkisovJETP97}. Another prominent effect is the 
generation of coherent structures such as electromagnetic solitons or vortices.
Numerical simulations 
(see e.g. \cite{sentokuPRL99,naumovaPRL01,esirkepovPRL02}) 
show that such structures are generated during the interaction with the laser 
pulse on a ultrafast (femtosecond) timescale, but they may lead to typical 
field structures which last for much longer times (e.g. ``post-solitons'')
\cite{naumovaPRL01} i.e. in the picosecond range, 
allowing for their experimental observation \cite{borghesiPRL02}. On such a 
scale the temporal evolution of such field structures must be studied including
the effects of the motion of the plasma ions. Stability and evolution of
coherent structures on the ion time scale has been studied theoretically
and numerically in several papers, for various regimes and dimensionality
\cite{sentokuPRL99,naumovaPRL01,esirkepovPRL02,farinaPRL01,lontanoPP03,lehmannPP06,weberPP05,ricondaPP06,liPP07}.

In this paper we report a theoretical study of nonlinear effects during
and after the propagation of a superintense laser pulse in an underdense,
longitudinally inhomogeneous plasma. The work was motivated
by experiments on laser
propagation in a low-density plasma where the dynamics of self-generated,
slowly varying electromagnetic fields was investigated using the 
proton diagnostic technique \cite{borghesiPPCF01}. In this paper we focus
on the simulation results and on their theoretical interpretation, while
a comparison with the experimental results will be reported elsewhere
\cite{karXXX07,liseykinaXXX07}.

\section{Simulation set-up}
The laser-plasma interaction simulations were performed using a 
particle-in-cell (PIC) code in 2D with Cartesian
geometry. Reduction to 2D was dictated by the need to address relatively long
spatial and temporal scales, close to the experimental ones. Moreover
(as it will be clear from the discussion of the results) 
during the interaction sharp gradients in the 
field and current patterns are generated. Thus, a reasonable resolution is 
mandatory to resolve such details, pushing the memory requirements in 3D
much beyond present-day supercomputing capabilities. Among the set of 2D
simulations that were performed for the present study, the largest ones
employed a $7750\times 2400$ grid, with spatial resolution 
$\Delta x=\Delta y=\lambda/10$ (where $\lambda$ is the laser pulse
wavelength) and 16 particles per cell for both electrons
and ions, requiring a total of 5000 CPU hours on 100 processors to simulate
more than 1500 laser periods of the interaction. 
The code is fully parallelized and the simulations were
performed at the CINECA supercomputing facility in Bologna (Italy).

In the following, lengths are given in units of $\lambda$, times in 
units of $T_L=\lambda/c=2\pi/\omega$, electric and magnetic fields in units of 
$E_0=m_e\omega c/e$, and densities in units of $n_c=m_e\omega^2/4\pi e^2$.
For $\lambda=1~\um$, $E_0=3.213 \times 10^{10}~\Vcm=107.1~\MG$ and 
$n_c=1.11 \times 10^{21}~\cm3$. The dimensionless parameter $a_L$, 
giving the peak field amplitude of the laser pulse normalized to $E_0$,
is related to the laser intensity $I$ and wavelength by
$a_L=0.85(I\lambda^2/10^{18}~\Wcmum2)^{1/2}$.

In all the 2D simulations reported below, the plasma is inhomogeneous
along the $x$ axis, i.e. in the direction of propagation of the laser
pulse. The electron density profile rises linearly from zero value at 
$x=25\lambda$ to the peak value $n_0=0.1n_c$ at $x=425\lambda$, and then
remains uniform.
The pulse duration $\tau_L$ was either $150$ or $300~T_L$, corresponding 
to $0.5$ and $1~\psec$, respectively, for $\lambda=1~\um$.

The laser pulse was $S$-polarized, i.e. the electric field of the laser pulse 
was in the $z$ direction perpendicular to the simulation plane. 
In the following we restrict the discussion to the $S$-polarization case which has some advantages for the data analysis 
and visualization (for instance, the space-charge field generated in 
radial ($y$) direction during self-channeling is separated by the 
electromagnetic field $E_z$ which is representative of the pulse evolution).
It is known, however, that at high intensity the details of nonlinear effects 
in pulse propagation depend on the polarization, leading to differences 
between the $S$- and $P$-polarization cases in 2D geometry and to asymmetry 
effects in 3D for what concerns self-focusing \cite{naumovaPP01} and also
to differences in the type and stability of solitons and vortices
\cite[and references therein]{bulanovRPP01,mourouRMP06}. 
A preliminary simulation performed for $P$-polarization showed slight, 
but no substantial differences for what concerns the early self-channeling 
evolution which we discuss in section~\ref{sec:self-channeling}. The 
discussion of the effect of different polarizations on the coherent 
structures generation and evolution is more involved and will be addressed in 
future work.

\section{Results}

To illustrate the variety of nonlinear effects observed in the simulation 
results, Figure~\ref{fig:ensemble} shows snapshots at $t=10^3 T_L$ of the 
ion density ($n_i$) and the electric field of the laser pulse ($E_z$)
over nearly the whole length of the plasma, for a simulation with 
$a_L=2.7$ and $\tau_L=330T_L.$ 
Figure \ref{fig:ensemble} contains most of the prominent features
we observed throughout the set of our simulations, which
may be summarized as follows. 

In the low-density region, the laser pulse bores a single 
charge-displacement channel, which in the higher density region 
breaks up into three main channels and a few of secondary, narrow filaments.
In the following (see section~\ref{sec:self-channeling})
we trace back the appearance of the ``trifurcated'' 
channel to the effects of radial ion acceleration which lead to the 
``breaking'' of the channel walls.

Different types of electromagnetic structures are observed in
regions of different density. In the lower density region
(approximately between $x=100$ and $x=150$ in Figure~\ref{fig:ensemble})
a pattern of fields with approximate axial symmetry is observed. 
The detailed analysis of the electric and magnetic fields, including an 
estimate of their characteristic frequency from the simulation 
(see section~\ref{sec:scavons}) shows that this type of structures
combines both features of low-frequency electromagnetic 
post-solitons or ``cavitons'' 
and steady current vortices. In the higher density region a number of 
slowly-evolving field structures, either appearing as ``solitary'' structures
or organized into patterns, are observed both outside the main low-density 
channels and inside the latter. There is some experimental indication of
the growth of regularly spaced field structures into the main channel 
\cite{liseykinaXXX07}.

\subsection{Ion and electric field dynamics following self-channeling}

\label{sec:self-channeling}
For intensities up to $a_L \simeq 2$, in the early stage of the interaction 
the laser pulse bores a regular charge-displacement channel in the 
inhomogeneous region of the plasma, i.e. at densities $n_e<0.1n_c$.
This is the case for the simulation of Figure~\ref{fig:self-channeling} 
($a_L=2$, $\tau_L=300T_L$, transverse width $r_L=4~\lambda$)
which shows a snapshot of the ion density $n_i$
and the electric field components $E_z$ and $E_y$ (results from this
simulation are also reported in \cite{karXXX07}). 
The laser pulse undergoes self-focusing as indicated both by the reduction of
its transverse radius to $\simeq 3\lambda$ and the increase of its
amplitude by a factor $\simeq 1.2.$

In the leading edge of the channel the transverse field $E_y$ is in the 
outward direction from the axis, indicating that the channel is positively 
charged due to the radial expulsion of electrons. In the trailing part of the 
pulse, the radial profile of $E_y$ changes qualitatively, as two ambipolar 
fronts appear on each side of the channel. On the inner side of the ambipolar 
fronts $E_y$ now points in the inward direction, i.e. towards the axis.
The onset of an ``inversion'' in the radial field has been noticed in 
experimental investigations of channel dynamics \cite{karXXX07}.

\subsection{One-dimensional modeling and the electric field ``echo''}

The dynamics leading to the evolution of the radial electric field 
can be studied in detail using an one-dimensional, electrostatic PIC model
where the laser pulse action is taken into account only via the ponderomotive
force. The model assumes a non-evolving radial profile of the laser pulse
and cylindrical symmetry, taking only the radial, cycle-averaged dynamics of
electron and ions into account. Details about the model and its results are 
reported elsewhere \cite{macchiXXX07}. Here we focus on the most prominent 
features of electric field dynamics. 

Figure~\ref{fig:1D} shows snapshots of the
radial electric field $E_r$ and the ion density $n_i$ at various times,
for a 1D simulation in the same regime of 2D electromagnetic runs.
Initially, the ponderomotive force $F_p$ pushes electron away from the axis, 
creating a back-holding space-charge field which is found to balance $F_p$ 
almost exactly. At the end of the pulse, when $F_p=0$, $E_r$ has almost 
vanished. However, $E_r$ appears back at a later time, with an ambipolar
profile very similar to that observed in the 2D simulations.
This ``echo'' effect originates from the ion dynamics of ions which are 
accelerated by the electric force $ZeE_r=ZF_p$ during the laser pulse.
The spatial profile of $F_p$ is such that the ions are focused towards
a very narrow region at the edge of the channel, producing a very sharp
spike of the ion density and leading to hydrodynamical breaking as
the fastest ions overturn the slowest ones. Looking at the profile 
of the ion density we observe that the latter may be said to ``break'' 
in literal meaning, as a secondary density spike moving outwards is formed.
The process is also accompanied by strong heating of electrons near the
breaking point, leading to the appearance of an ambipolar sheath field
around the density spike. The negative field is strong enough to slow down
and invert the velocity of the slowest ions, which are directed back to the 
axis where they are found to form a local density maximum at later times.

\subsection{Laser beam breakup}

A simple analytical model shows that the time required for the ions
in the channel to reach the ``breaking'' point is proportional to 
the channel radius and inversely proportional to the laser field amplitude
\cite{macchiXXX07}. For high intensities, the ``breaking'' effect 
due to ion acceleration may occur early during the laser pulse, 
i.e. when the electromagnetic energy density inside the channel is 
very high, and cause a fast, strong variation of the density at the edge 
of the channel. In turn, this may affect the propagation of the laser pulse,
similarly to what would happen in a wave guide where a sudden ``leak'' 
in its walls occurs. A possible signature of this effect is the appearance
of two secondary beams, propagating in oblique direction, and originating
near the point where the breaking of the channel walls occurs, as can be
observed in Figure~\ref{fig:breakup}.

From the ``leaking waveguide'' picture we roughly estimate these secondary
beams to propagate at an angle $\theta$ with respect to the axis given 
by $\tan\theta \simeq k_y/k_x$, where $k_y \simeq \pi/d$ is the transverse 
wavevector of the guided mode, $d$ is the local channel diameter, and 
$k_x \simeq \sqrt{\omega^2/c^2-k_y^2}$. In this estimate the pulse 
in channel is modelized as a TE mode of lowest order in a square guide.
From the simulation result we get $\tan\theta \simeq 0.065$, while
being $d \simeq 7\lambda$ we obtain
$k_y/k_x \simeq (\pi/7\lambda)/(2\pi/\lambda) = 0.071$.

\subsection{Slowly-varying electromagnetic structures}

\label{sec:scavons}
As already noticed in Figure~\ref{fig:ensemble} an impressive number of 
localized, slowly varying structures is generated in the interaction. 
In the denser plasma region, the several small-scale structures 
whose most evident signature is a strong depression in the plasma density 
are likely to be rather similar to the so-called post-solitons 
\cite{naumovaPRL01,borghesiPRL02}, having zero propagation velocity and
slowly expanding due to ion acceleration driven by the internal radiation
pressure. They may be described as small cavities trapping electromagnetic 
radiation whose frequency is less than the plasma frequency of the 
surrounding plasma (hence they may be also appropriately named as 
electromagnetic ``cavitons''). We notice that we do not observe a drift
of such structures towards the low-density region. This difference from
the observations of Ref.\cite{sentokuPRL99} might be ascribed
to the smoother electron gradient in our case.

The regular structures, forming an axially symmetrical row, observed in the 
low density region near the plasma boundary (far left side in 
Figure~\ref{fig:ensemble}) have indeed features which are similar 
both to electromagnetic cavitons and magnetic vortices. This ``dual'' 
nature can be observed in Figure~\ref{fig:scavons}, which shows the components 
of the fields $E_z$ and $B_z$ perpendicular to the simulation plane as a 
contour plot and the components in the $(x,y)$ plane as a vector 
plot.
By analyzing the frequency spectrum of the fields inside the density 
depression, we find that the fields $E_z$, $B_x$ and $B_y$ are oscillating at 
a frequency of approximately $0.1\omega$, lower than the local value of the 
plasma frequency (for unperturbed plasma) $\omega_p \simeq 0.15\omega$.
Qualitatively, the oscillating fields are similar to those of the 
lowest TM resonant mode in a cylindrical cavity.

The frequency analysis of $E_x$, $E_y$ and $B_z$ shows that these field 
components are quasi-static, their spectrum being peaked around zero
frequency. The electric field components $E_x$ and $E_y$ are in radial
direction with respect to the axis of the structure, as it is expected for
a cavity expanding under the action of the radiation pressure of the trapped
radiation. The static magnetic field component $B_z$ is associated to 
current rings flowing around the axis of the structure. 

Apart from being associated to ``post-soliton''-like structures, the 
fact that the magnetic vortices form a symmetrical row and are localized 
near the boundary of the channel make them different from those observed 
in the wake of a much shorter laser pulse, for which the creation of a 
low-density channel does not occur, and which seem to form an antisymmetrical 
row \cite{askaryanPPCF97,bulanovPRL96}. It is nevertheless possible that
the current filamentation instability discussed in Ref.\cite{bulanovPRL96} 
plays a role in vortex formation also
in the present case. In the early stage we observe a strong electron current 
in the main channel and two narrow return current sheets just outside the
channel boundaries; later, the current layers seem to bend locally
forming vortices around magnetic field maxima.

The axial symmetry of these particular structures suggests that in 
``realistic'' 3D geometry they may have a toroidal or ``donut'' shape. To get 
an impression of such a 3D structure one should imagine to rotate the field 
patterns of the 2D simulations around the $x$ axis. This particular type of 
coherent structure would be characterized by azimuthal components of 
${\bf E}$ (oscillating) and ${\bf B}$ (quasi-static) directed along 
the torus circumference, a solenoidal and oscillating magnetic field coiled up 
round the torus, and by an electrostatic fields component perpendicular to the 
torus surface. The 3D soliton discussed in Ref.\cite{esirkepovPRL02}
has a toroidal magnetic field and a poloidal magnetic field; however,
in our case we have no clear indication of the charge oscillations 
inside the solitons observed in Ref.\cite{esirkepovPRL02}.

Inside the main and secondary low-density channels generated in the denser
region of the plasma, the growth of field patterns which are 
less regular than those of Figure~\ref{fig:scavons}, but qualitatively similar
can be observed. Figure~\ref{fig:scavons2} give details of their evolution.
We observe a tendency of this type of structures to grow inside the channels
and to be correlated with rippling and bending of the channel walls. 
Theoretical work will be required
to address the physics of formation of such structure patterns.

\section{Conclusions}

The main results emerging from the series of 2D PIC simulations reported
in the present paper may be summarized as follows. Ion acceleration due
to the space-charge field in the channel drilled by the laser pulse 
leads to hydrodynamical breaking of the plasma profile at the channel walls.
Two side effects of the ion-driven ``breaking'' have been identified:
a change in the radial profile of the electrostatic field (including a sort
of ``echo'' effect for pulses shorter than the breaking time) and a breakup
of ``long'' laser pulses due to the sudden ``leak'' generated in the channel
walls. The evolution of coherent, slowly-varying field structures has been
monitored in time up to thousand of laser cycles, corresponding to several
picoseconds in ``real'' experiments. Patterns of multi-peak structures 
appear inside low-density channels, and the formation of structures having 
both oscillating and static field components with an hybrid soliton-vortex
nature has been observed. These results, and the perspective of experimental
investigations of such field patterns, support the view of relativistic
``laser plasmas'' as environments showing an high degree of self-organization
and a wealth of coherent structures, which are thus of great interest for 
the physics of nonlinear systems.

\ack
This work has been supported by the Royal Society (UK) via a Joint Project, by
the Ministry of University and Research (Italy) via a PRIN project,
and by CNR-INFM and CINECA (Italy) through the super-computing 
initiative. Valuable discussions with S. V. Bulanov and F. Pegoraro and
the help of E. Echkina in the analysis of some simulations are warmly
acknowledged.

\section*{References}

\bibliographystyle{unsrt}
\bibliography{Paper_EPS_07_macchi}

\newpage

\begin{figure}
\begin{center}
\caption{Contours of the electric field of the laser pulse $E_z$ and the 
ion density $n_i$ from a 2D PIC simulation, at $t=1000T_L$, showing the 
self-channeling and filamentation of the laser pulse and the generation 
of isolated solitary structures and of field patterns. The laser pulse 
propagates from left to right. In the $E_z$ frame, the contour levels 
in the leftmost region ($50<x<250$) has been rescaled by a factor of 3 to
show the presence of field structures with relatively low amplitudes.
The laser pulse parameters are $a_L=2.7$, $r_L=8\lambda$, and 
$\tau_L=330T_L.$ 
\label{fig:ensemble}}
\end{center}
\end{figure}

\begin{figure}
\begin{center}
\end{center}
\caption{Simulations results addressing electric field dynamics 
following self-channeling,
showing the transition in the radial field profile \cite{karXXX07}.
Left column: 2D PIC results.
Top frame: ion density ($n_{i}$) and electric field components
($E_{z}$ and $E_{y}$) at $t=600T_L$. 
Bottom frame: lineout of $E_{y}$ (blue) and $n_{i}$ (red) along
the $y$-axis at two different $x$-positions. 
Parameters are $a_L=2$, $\tau_L=300T_L$, $r_L=4~\lambda$.
Right column: snapshots at various times 
of radial electric field $E_r$ (blue, thick line)
and ion density $n_i$ (red, dash-dotted line), 
and the phase space distributions
of ions $f_i(r,p_r)$ and electrons $f_e(r,p_r)$
from 1D simulations using a ponderomotive, 
electrostatic model \cite{macchiXXX07}.
Parameters are $a_L=2.7$, $n_e/n_c=0.01$, $r_L=7.5\lambda$,
$\tau_{L}=330T_L$.
\label{fig:self-channeling}\label{fig:1D}}
\end{figure}

\begin{figure}
\begin{center}
\end{center}
\caption{Evolution of the laser field $E_z$ at different times showing the 
breakup of the laser pulse into three main beams. 
The two secondary beams propagating in oblique
direction originate from near the location of the ``breaking'' of the 
channel walls.
The laser pulse parameters are $a_L=2.7$, $r_L=8\lambda$, and 
$\tau_L=150T_L.$ 
\label{fig:breakup}}
\end{figure}

\begin{figure}
\begin{center}
\end{center}
\caption{(Anti-)symmetrical row of slowly-varying structures 
in the low density region of the plasma at $t=625T_L$. 
The left column shows the fields $E_z$ (contour plot) and
${\bf B}_x+{\bf B}_y$ (vector plot) oscillating at a frequency 
$\simeq 0.1\omega$. 
The right column shows the quasi-static fields $B_z$ (contour plot) and
${\bf E}_x+{\bf E}_y$ (vector plot). 
The laser pulse parameters are $a_L=2.7$, $r_L=8\lambda$, and 
$\tau_L=150T_L.$ 
\label{fig:scavons}}
\end{figure}

\begin{figure}
\begin{center}
\end{center}
\caption{Detail of the evolution of field structures in the denser region of
the plasma, from the simulation of Figure~\ref{fig:ensemble}.
The laser pulse parameters are $a_L=2.7$, $r_L=8\lambda$, and 
$\tau_L=300T_L.$ 
\label{fig:scavons2}}
\end{figure}

\end{document}